\documentstyle[12pt]{article}
\makeatletter
\@addtoreset{equation}{subsection}
\makeatother

\begin{document}
\begin{center}
{\Large \bf
Indeterministic Quantum Gravity and Cosmology} \\[0.5cm]
{\large\bf VI. Predynamical Geometry of Spacetime
Manifold, Supplementary Conditions for Metric, and CPT} \\[1.5cm]
{\bf Vladimir S.~MASHKEVICH}\footnote {E-mail:
mash@phys.unit.no}  \\[1.4cm]
{\it Institute of Physics, National academy
of sciences of Ukraine \\
252028 Kiev, Ukraine} \\[1.4cm]
\vskip 1cm

{\large \bf Abstract}
\end{center}

This paper is a continuation of the papers [1-5]. The
introduction of a prior, i.e., predynamical global geometry
of spacetime manifold is substantiated,
and the geometry is specified. The
manifold is an infinite four-cylinder, or tube in the
five-dimensional Euclidean space, the orthogonal section of the
cylinder being the unit three-sphere. Supplementary conditions
for metric are introduced geometrically, coordinate-independently,
as opposed to coordinate conditions. Parity and time-reversal
transformations are extended to the manifold specified. $PT$
is a rotation through $\pi$ about an axis orthogonal to the
cylinder axis. $CPT$ invariance is discussed.

\newpage

\hspace*{6 cm}
\begin{minipage} [b] {9 cm}
One geometry cannot be more true than another; it can only
be more {\it convenient}. Geometry is not true, it is
advantageous.
\end{minipage}
\begin{flushright}
Robert Pirsig \vspace*{0.8 cm}
\end{flushright}

\begin{flushleft}
\hspace*{0.5 cm} {\Large \bf Introduction}
\end{flushleft}

In general relativity (GR), a spacetime manifold $M$ is a
four-dimensional, real, Hausdorff, $C^\infty$, and paracompact
manifold, which, however, possesses no prior, i.e., predynamical
geometry. To emphasize the last issue, we quote [6]:
`Mathematics was not sufficiently refined in 1917 to cleave
apart the demands for ``no prior geometry'' and for a ``geometric,
coordinate-independent formulation of physics''. Einstein
described both demands by a single phrase, ``general covariance''.
The ``no-prior-geometry'' demand actually fathered general
relativity, but by doing so anonymously, disguised as ``general
covariance'', it also fathered half a century of confusion'.

Another characteristic feature of GR is locality, `This is
surely Einstein's concept that all physics takes place by ``local
action'''[6].

On the other hand, in the theory being developed in this series
of papers, it is a prior global geometry of spacetime manifold
that is of considerable importance. The first aim of this paper
is to substantiate the introduction of this geometry and to
specify
the latter. (Since the role of the global geometric structure of
the spacetime manifold is essential for the theory, we name it
indeterministic quantum gravity and cosmology.)

A motivation for introducing the global structure is related to
dynamical incompleteness of the Einstein equation and coordinate
conditions. This issue was considered in [3,4]. The second aim
of this paper is to continue the consideration.

The global structure allows for extending $PT$ transformation
and $CPT$ invariance to curved spacetime, which is the third aim
of this paper.

The main results are as follows. The spacetime manifold is an
infinite four-cylinder, or tube in the five-dimensional Euclidean
space, the orthogonal section of the cylinder being the unit
three-sphere. Supplementary conditions for metric $g$ are
introduced geometrically, coordinate-independently;
$g=dt\otimes dt-h_t$ where $h_t$ is a time-dependent Riemannian
metric on the sphere. $PT$ is equivalent to a rotation through
$\pi$ about an axis orthogonal to the cylinder axis.

\section{Motivation for predynamical global geometry}

\subsection{Dynamical incompleteness of the Einstein equation}

A motivation for introducing a global predynamical structure
of spacetime manifold in [3] was based on the dynamical
incompleteness of the Einstein equation. Let us return to
this issue. In the Einstein equation
\begin{equation}
G=T
\label{1.1}
\end{equation}
($G$ is the Einstein tensor and $T$ is the energy-momentum
tensor), as it is well known, there are only six dynamical
equations
\begin{equation}
G^{ij}=T^{ij},\quad i,j=1,2,3,
\label{1.2}
\end{equation}
for six metric components $g_{ij}$, whereas four equations
\begin{equation}
G^{0\mu}=T^{0\mu},\quad \mu=0,1,2,3,
\label{1.3}
\end{equation}
are constraints on initial data, so that four components
$g_{0\mu}$ are dynamically undetermined. The
conventional way for circumventing the dynamical incompleteness
consists in introducing gauge, or coordinate conditions [7,8].
The reasoning behind those conditions is as follows. Let $M$
be a spacetime manifold,
\begin{equation}
{\cal F}(M)=\bigcup\limits_{p\in M}{\cal F}_p,\qquad {\cal F}
=(g,{\cal M})
\label{1.4}
\end{equation}
where $\cal M$ is a set of matter fields, and
\begin{equation}
\mu:M\to M
\label{1.5}
\end{equation}
be a differentiable transformation. The set $\mu_*\cal F$
is physically
equivalent to $\cal F$. The transformation $\mu$ involves four
gauge functions, which may be used to remove the indeterminacy
of the $g_{0\mu}$'s. We stress that, as it is known, the gauge
conditions cannot be formulated in covariant form [8].

We argue against that approach to the incompleteness of dynamics
as follows. There are two alternative approaches:
Spacetime manifold is a
set of events $E$ [9,10,6]; spacetime manifold is an abstract
manifold $M$ [7].

In the first case, we have ${\cal F}(E)$, so that $\mu_*\cal F$
is not physically equivalent to $\cal F$ since at least $g_e,
e\in E$, may be measured [7] and $(\mu_*g)_e\ne g_e$. Let us try
to remedy the situation. Let
\begin{equation}
\nu:E\to M
\label{1.6}
\end{equation}
be a diffeomorphism, where $M$ is an abstract manifold, then
\begin{equation}
\nu_*{\cal F}(E)={\cal F}^M(M).
\label{1.7}
\end{equation}
Now we introduce $\mu$ (\ref{1.5}) and obtain
\begin{equation}
\overline{\cal F}^M=\mu_*{\cal F}^M,
\label{1.8}
\end{equation}
which is physically equivalent to ${\cal F}^M$. In $\overline
{\cal F}^M$ the indeterminacy of the $g_{0\mu}$'s may be removed,
so that $\overline{\cal F}^M$ may be considered to be known. But
we cannot find
\begin{equation}
{\cal F}=\nu^*\mu^*\overline{\cal F}^M
\label{1.9}
\end{equation}
since $\mu$ is not known.

In the second case, we invoke the geometric principle, i.e., the
demand for a geometric, coordinate-independent formulation of
physics [6]. Since the principle cannot be realized locally in a
covariant form, it should be realized globally---in the form of
predynamical global geometry and related supplementary (not
coordinate!) conditions for metric.

At the conclusion of this subsection, we note that any
coordinate conditions imply a single global coordinate
system, which, in general, does not exist.

\subsection{Cauchy problem vicious circle}

An initial-value problem in dynamics is the Cauchy problem.
Metric is obtained from a solution to this problem. The
Cauchy problem for a given spacetime manifold demands a
given family of Cauchy surfaces. But a Cauchy surface is
determined by metric. Thus we come to a vicious circle:
Metric implies the Cauchy problem, the Cauchy problem
implies a Cauchy surface, the Cauchy surface implies metric,

\hspace{6,2cm} metric

\hspace{5,5cm} $\nearrow$\hspace{1,5cm}  $\searrow$

\hspace{4cm} Cauchy surface $\gets$ Cauchy problem

\noindent where arrow stands for implies.

To break the vicious Cauchy circle, it is necessary to
introduce a family of Cauchy surfaces, which may be done on
the basis of a predynamical global geometry.

\subsection{Quantum jumps}

To construct an indeterministic theory, it is necessary to
incorporate quantum jumps into dynamics. The jumps imply the
existence of a preferred time, i.e., cosmic time, which, in
its turn, implies a specific global structure of spacetime
manifold.

\section{Maximally symmetric cylindrical manifold}

\subsection{Spacetime as a set}

To be a mathematical object, spacetime should first of all
be defined as a set. We invoke Georg Cantor [11,12]: `I call
a manifold (a totality, a set) of elements which belong to
some conceptual sphere well-defined, if on the basis of its
definition and as a consequence of the logical principle of
excluded middle it must be seen as {\it internally determined
both} whether some object belonging to the same conceptual
sphere belongs to the imagined manifold as an object or not,
{\it as well as} whether two objects belonging to the set are
equal to one another or not, despite formal differences in the
way they are given'. In the case of spacetime manifold $M$,
a straightforward definition may be based on the Whitney
embedding theorem: The problem of the identification of points
of $M$ is naturally solved by considering $M$ as a subset of
$R^p$, since points of $R^p$ are identified. Here $R^p$ is
the Euclidean (or arithmetical) space with
\begin{equation}
4\le p\le 9\;(=2\cdot 4+1).
\label{2.1}
\end{equation}
In fact, there is no other natural way for solving the problem.

In the indeterministic cosmology being developed in this series
of papers, the universe should be spatially finite---otherwise
the cosmic energy determinacy principle [1] would make no sense.
Assuming, in addition, that the universe is spatially closed we
have
\begin{equation}
5\le p\le 9.
\label{2.2}
\end{equation}
We adopt the simplest possibility, $p=5$. Thus
\begin{equation}
M\subset R^5.
\label{2.3}
\end{equation}

\subsection{Trivial bundle. Cylindrical manifold}

The existence of cosmic time implies a fibration of spacetime
manifold:
\begin{equation}
\pi:M\stackrel{S}{\longrightarrow}T,
\label{2.4}
\end{equation}
so that $M$ is a bundle space, a base space $T$ is cosmic time,
a standard fibre $S$ is cosmic space, and $\pi$ is the projection.
Cosmic time is an open interval in the real axis,
\begin{equation}
T=(a,b)\subset R.
\label{2.5}
\end{equation}
The simplest way of forming a fibre bundle (\ref{2.4}) is to
take the product
\begin{equation}
M=T\times S,
\label{2.6}
\end{equation}
i.e., the trivial bundle. Thus $M$ is a four-dimensional
cylinder, or tube in $R^5$.

\subsection{Maximal symmetry. Spherical space manifold}

From simplicity desideratum, we impose the maximal symmetry
conditions on the manifold $M$ (\ref{2.6}): In the Euclidean
metric of $R^5$
\begin{equation}
T\subset R\perp R^4\supset S
\label{2.7}
\end{equation}
and
\begin{equation}
S={\cal S}^3
\label{2.8}
\end{equation}
where ${\cal S}^3$ is the unit 3-sphere, so that $M$ is a
right 4-cylinder whose orthogonal section is ${\cal S}^3$.

Note that $M$ corresponds to the manifold in the
Robertson-Walker spacetime of positive spatial curvature.

\section{Supplementary conditions for metric}

\subsection{Tangent space}

By eq.(\ref{2.6}) the tangent space at a point $p\in M$ is
\begin{equation}
M_p=T_p\oplus S_p.
\label{3.1}
\end{equation}
In the metric induced on $M$ by the Euclidean metric of
$R^5$, we have
\begin{equation}
T_p\perp S_p\quad {\rm in\; the\; Euclidean\; metric.}
\label{3.2}
\end{equation}

\subsection{Orthogonality condition}

As a key condition on the metric $g$, we adopt
the following: $g$ should respect the orthogonality
condition (\ref{3.2}),
\begin{equation}
T_p\perp S_p\quad {\rm in\;the\;metric}\;g.
\label{3.3}
\end{equation}
This condition is introduced from symmetry, i.e., once again,
simplicity desideratum. It should be particularly emphasized
that the condition is formulated geometrically,
coordinate-independently.

Eq.(\ref{3.3}) implies
\begin{equation}
g=g_T+g_S,\qquad g_T=g_{00},\qquad g_S=g_{ij}.
\label{3.4}
\end{equation}

Using a standard scaling for time and taking into account the
Lorentzian character of the metric, we obtain
\begin{equation}
g=dt\otimes dt-h_t,
\label{3.5}
\end{equation}
where $h_t$ is a time-dependent Riemannian metric on
${\cal S}^3$. In the coordinate form, eq.(\ref{3.5}) reads
\begin{equation}
ds^2=dt^2-h_{ij\,t}dx^idx^j,\quad dx^0=dt.
\label{3.6}
\end{equation}
Thus the dynamics of spacetime is described by six metric
components $g_{ij}=-h_{ij}$, which corresponds to the six
dynamical equations (\ref{1.2}).

\section{CPT and the eternal universe}

The aim of this section is to extend parity and time-reversal
transformation to a curved spacetime with the manifold $M$
given by eqs.(\ref{2.6}),(\ref{2.5}),(\ref{2.8}),
$$
M=(a,b)\times{\cal S}^3,
\eqno{(4.0.1)}
$$
so as to obtain the possibility of $CPT$ invariance.

\subsection{Parity}

We begin with introducing parity $P$ for the unit sphere.
For the sake of clarity of presentation, we consider 1-
and 3-sphere in $R^2$ and $R^4$ respectively.

1-sphere:

\begin{eqnarray}
x^2+y^2=1,\nonumber\\
y=\sin\varphi\nonumber\\
x=\cos\varphi,\nonumber\\
-\pi\le\varphi\le\pi;\nonumber\\
P:\varphi\to-\varphi,\nonumber\\
y\to-y,\; x\to x;
\label{4.2}
\end{eqnarray}

3-sphere:

\begin{eqnarray}
x^2+y^2+z^2+u^2=1,\nonumber\\
u=\sin\chi\nonumber\\
z=\cos\chi\cdot\sin\vartheta\nonumber\\
y=\cos\chi\cdot\cos\vartheta\cdot\sin\varphi\nonumber\\
x=\cos\chi\cdot\cos\vartheta\cdot\cos\varphi,\nonumber\\
-\pi/2\le\chi\le\pi/2,\;-\pi/2\le\vartheta\le\pi/2,\;
-\pi\le\varphi\le\pi;\nonumber\\
P:\chi\to-\chi,\;\vartheta\to-\vartheta,\;\varphi\to-\varphi,
\nonumber\\
u\to-u,\;z\to-z,\;y\to-y,\;x\to x.
\label{4.4}
\end{eqnarray}
Thus for the $(2n+1)$-sphere, $n=0,1,...,$
the parity transformation $P$ is
given by the
reflections in $2n+1$ hyperplanes which intersect along
a diameter and
are mutually orthogonal, or, to put this another way, by the
inversion with respect to the diameter. In eqs.(\ref{4.2}),
(\ref{4.4}), the diameter is along the $x$ axis.

For the $(2n+1)$-sphere in $R^{2n+3}$, $P$ is equivalent to a
rotation through $\pi$ about the axis $x$, i.e., through
$\pi$ in $n+1\;\; 2$-planes orthogonal to the axis $x$ and to each
other.

\subsection{Time reversal and the eternal universe}

To introduce time-reversal transformation,
\begin{equation}
T:t\to-t,
\label{4.5}
\end{equation}
we should adopt in eqs.(\ref{2.5}),(4.0.1)
\begin{equation}
(a,b)=R,
\label{4.6}
\end{equation}
so that $M$ becomes an infinite cylinder,
\begin{equation}
M=R\times{\cal S}^3.
\label{4.7}
\end{equation}
For the infinite cylindrical manifold, time reversal is the
reflection in a hyperplane orthogonal to the cylinder axis.

The infinite time interval (\ref{4.6}) implies the universe
to be eternal. One of the aspects of such a universe (namely,
related to the oscillating model) was discussed in [1].

\subsection{PT}

For $2m$-cylinder in $R^{2m+1}$ (in our case $m=2$), $PT$ is
equivalent to the rotation through $\pi$ about an axis (in our
case the $x$ axis) orthogonal to the cylinder axis, so that we
introduce the notation
\begin{equation}
R=PT
\label{4.8}
\end{equation}

\subsection{CPT}

Now it is possible to consider $CPT$ invariance conditions.
The dynamics of indeterministic cosmology is described by
equations of motion between quantum jumps [2] and relations
for the jumps [5]. The equations of motion are (II.4.1,4.2):
\begin{equation}
G_S=(\Psi,T_S\Psi),
\label{4.9}
\end{equation}
\begin{equation}
H\Psi=\varepsilon\Psi.
\label{4.10}
\end{equation}
The jump relations reduce, in the final analysis, to the
between-jump dynamics and standard probabilities
\begin{equation}
w=\mid(\Psi_2,\Psi_1)\mid^2.
\label{4.11}
\end{equation}
We have
\begin{equation}
G=G(g,g',g''),\quad T=T(g,g'),\quad H=H[g,g']
\label{4.12}
\end{equation}
where prime denotes derivatives with respect to $x^\mu,
\mu=0,1,2,3$.

$CPT$ transformation is

\begin{equation}
g\to Rg,\qquad \Psi\to V\Psi
\label{4.13}
\end{equation}
where
\begin{equation}
V=U_CU_PV_T,
\label{4.14}
\end{equation}
the $U$'s are unitary and the $V$'s antiunitary. Transformed
eqs.(\ref{4.9})-(\ref{4.11}) are
\begin{equation}
G_S(Rg,Rg',Rg'')=(V\Psi,T_S(Rg,Rg')V\Psi),
\label{4.15}
\end{equation}
\begin{equation}
H[Rg,Rg']V\Psi=\varepsilon V\Psi,
\label{4.16}
\end{equation}
\begin{equation}
w=\mid(V\Psi_2,V\Psi_1)\mid^2.
\label{4.17}
\end{equation}

Since
\begin{equation}
(V\Psi_2,V\Psi_1)=(\Psi_1,\Psi_2),
\label{4.18}
\end{equation}
eq.(\ref{4.17}) reduces to eq.(\ref{4.11}).

Since $G$ is quadratic in differentiation operators,
\begin{equation}
G_{S\,p}(Rg,Rg',Rg'')=G_{S\,Rp}(g,g',g'')
\label{4.20}
\end{equation}
holds. We have
\begin{equation}
(V\Psi,T_S(Rg,Rg')V\Psi)=(V\Psi,VV^{-1}T_S(Rg,Rg')V\Psi).
\label{4.21}
\end{equation}
Let
\begin{equation}
V^{-1}T_{S\,p}(Rg,Rg')V=T_{S\,Rp}(g,g')
\label{4.22}
\end{equation}
be fulfilled, then
\begin{equation}
(V\Psi,T_S(Rg,Rg')V\Psi)=(V\Psi,VT_{S\,Rp}(g,g')\Psi)
=(\Psi,T_{S\,Rp}(g,g')\Psi)^*=(\Psi,T_{S\,Rp}(g,g')\Psi),
\label{4.23}
\end{equation}
so that eq.(\ref{4.15}) reduces to eq.(\ref{4.9}).

We have from (\ref{4.16})
\begin{equation}
V^{-1}H[Rg,Rg']V\Psi=\varepsilon \Psi.
\label{4.24}
\end{equation}
Let
\begin{equation}
V^{-1}H[Rg,Rg']V=H[g,g']
\label{4.25}
\end{equation}
be fulfilled, then eq.(\ref{4.16}) reduces to eq.(\ref{4.10}).

Thus the $CPT$ invariance conditions are eqs.(\ref{4.22}),
(\ref{4.25}).

\section{Quantum jumps and cosmology}

In conclusion, note the following. In the framework of
special relativity (SR), i.e., without taking into account
gravity, it is impossible to construct a dynamics of
quantum jumps: There is no preferred time in the structure
of Minkowskian spacetime. Both gravity and quantum jumps are
not consistent with SR, which links them. Both cosmology and
quantum jumps demand a preferred time, which links them. A
realization of that link is at the heart of indeterministic
cosmology.

\section*{Acknowledgment}

I would like to thank Stefan V. Mashkevich for helpful
discussions.

\end{document}